\documentclass[
  aip,
  jap,
  superscriptaddress,
  reprint,
  floatfix,
]{revtex4-1}

\usepackage{graphicx}
\usepackage{dcolumn}
\usepackage{amsmath,amssymb,bm}

\usepackage[utf8]{inputenc}
\usepackage[T1]{fontenc}
\usepackage{mathptmx}
\usepackage{etoolbox}

\usepackage{mathtools}
\usepackage[separate-uncertainty=true, parse-numbers=false]{siunitx}

\graphicspath{{./figure/}} 

\begin{document}

\title{Composition-Based Machine Learning for Screening Superconducting Ternary Hydrides from a Curated Dataset}

\author{Kazuaki Tokuyama}
\affiliation{Department of Chemical Engineering, Kyoto University,
	Kyoto 615-8510, Japan}

\author{Souta Miyamoto}
\affiliation{Department of Chemical Engineering, Kyoto University,
	Kyoto 615-8510, Japan}
\affiliation{School of Chemical Science and Technology, Kyoto University, Nishikyo, Kyoto 615-8510, Japan}

\author{Taichi Masuda}
\affiliation{Department of Chemical Engineering, Kyoto University,
	Kyoto 615-8510, Japan}

\author{Katsuaki Tanabe}
\thanks{Corresponding author.\\
	\href{mailto:tanabe@cheme.kyoto-u.ac.jp}{tanabe@cheme.kyoto-u.ac.jp}}
\affiliation{Department of Chemical Engineering, Kyoto University,
	Kyoto 615-8510, Japan}
\affiliation{School of Chemical Science and Technology, Kyoto University, Nishikyo, Kyoto 615-8510, Japan}
\affiliation{Kyoto MPI Inc., Sakyo, Kyoto 606-8501, Japan}

\date{\today}

\begin{abstract}
We present an ensemble machine-learning approach for composition-based, structure-agnostic screening of candidate superconductors among ternary hydrides under high pressure.
Hydrogen-rich hydrides are known to exhibit high superconducting transition temperatures, and ternary or multinary hydrides can stabilize superconducting phases at reduced pressures through chemical compression.
To systematically explore this vast compositional space, we construct an ensemble of 30 XGBoost regression models trained on a curated dataset of approximately 2000 binary and ternary hydride entries.
The model ensemble is used to screen a broad set of A–B–H compositions at pressures of 100, 200, and 300 GPa, with screening outcomes evaluated statistically based on prediction consistency across ensemble members.
This analysis highlights several high-scoring compositional systems, including Ca–Ti–H, Li–K–H, and Na–Mg–H, which were not explicitly included in the training dataset.
In addition, feature-importance analysis indicates that elemental properties such as ionization energy and atomic radius contribute significantly to the learned composition-level trends in superconducting transition temperature.
Overall, these results demonstrate the utility of ensemble-based machine learning as a primary screening tool for identifying promising regions of chemical space in superconducting hydrides.
\end{abstract}

\maketitle

\section{Introduction}
\label{sec:intro}

The search for superconductors operating near room temperature has remained a significant quest. 
Recently, hydrogen-rich materials have emerged as promising candidates for high superconducting transition temperatures \(T_\mathrm{c}\), inspired by Ashcroft’s theory on metallic hydrogen and hydride superconductivity\cite{Ashcroft1968-ds,Ashcroft2004-pg}. 
The concept has been validated through experimental progress, the discovery of sulfur hydride (H$_3$S) exhibiting superconductivity at 203 K under 150 GPa and lanthanum hydride (LaH$_{10}$) at 250 K under 170 GPa\cite{Drozdov2015-zz, Somayazulu2019-wt, Drozdov2019-jj}. 
Such experimental achievements have been supported by theoretical and computational methods, especially density functional theory (DFT) combined with Eliashberg theory of electron-phonon interactions\cite{Duan2014-fu,Liu2017-qz}.

Despite these developments, the utility of hydrides remains constrained due to the extremely high pressures. 
Recent theoretical and computational efforts expand their scope from binary to ternary or even multinary hydrides\cite{Zhao2024-af,Sun2024-qd}, exploiting chemical compression, thereby stabilizing superconducting phases at reduced pressures. 
Extensive experimental and theoretical studies have emerged for ternary hydride systems\cite{Wei2023-bp, Gao2024-ko, Gao2024-es, Wrona2024-gx, Liu2024-dd, He2024-ei, Huang2024-zw, Cerqueira2024-fp, Cerqueira2024-gb, Tian2024-mj, Gao2023-mz, Di-Cataldo2020-zo, Song2024-wc, Durajski2021-sr, Jiang2021-ug, Lv2020-ab, Han2019-xd, Hou2022-uv, Niu2023-rt, Guan2021-bw, Xie2019-cc, Shao2019-le, Zhao2023-yy, Song2022-kk, Shi2021-pg, Liang2019-qg, Ma2017-fb, Sukmas2020-zu, An2023-kq, Sun2022-er, Du2022-np, Hai2023-fq, Sun2022-yb, Wang2022-oy, Hu2022-vd, Liao2022-eb, Du2021-pk, Cui2020-vl, Hai2022-km, Sun2020-gy, Huo2023-hh, Wan2022-ja, Jiang2022-hg, Li2022-zx, Gao2021-bx, Lucrezi2022-xm, Di-Cataldo2021-hg, Liang2021-ye, Sun2022-bb, Jiang2024-uf, Sun2019-dd, Chen2023-bf, Semenok2021-wc, Song2023-ty, Zhang2022-wr, Gu2017-ko}. 
Notable examples, such as La–Be–H\cite{Zhang2022-wr} and Y–Mg–H\cite{Song2022-kk}, involve combinations of light elements, alkali metals, alkaline earth metals, and early transition metals. 
Exploring vast chemical spaces through experiments and first-principles calculations requires prohibitive resources. 

To overcome this limitation, machine learning (ML) methods have increasingly been utilized to navigate and screen chemical spaces, accelerating identifications of potential superconductors\cite{Stanev2018-dn,Konno2021-vg,Xie2019-qn,Xie2022-ld,Court2020-ij,Hutcheon2020-dk, Jiang2025-lr}.
For examples of general superconductors, Stanev \textit{et al.}\cite{Stanev2018-dn} introduced a framework combining a coarse classifier and regression models of cuprate, iron-based and low-\(T_\mathrm{c}\) compounds \(T_\mathrm{c}\), trained on the SuperCon dataset\cite{materials-a}.
Konno \textit{et al.}\cite{Konno2021-vg} developed a highly accurate ML model, incorporating periodic table features, achieving R\(^2=0.92\) on the SuperCon dataset.
Xie \textit{et al.}\cite{Xie2019-qn,Xie2022-ld} derived symbolic expressions estimating \(T_\mathrm{c}\), effectively serving as surrogates for the analytical Allen and Dynes formula\cite{AllenDynes1975-zz}, using a sparse regression method.
Courts and Cole\cite{Court2020-ij} applied natural language processing techniques to construct phase diagrams for superconducting and magnetic materials.
Notably, hydride-focused ML studies have become possible by the gradual accumulation of computational data, obtained from density functional perturbation theory.
Hutcheon \textit{et al.}\cite{Hutcheon2020-dk} investigated binary hydrides, predicting their high \(T_\mathrm{c}\) values using a neural network model.
More recently, Jiang \textit{et al.}\cite{Jiang2025-lr} developed a stability-oriented screening framework for ternary hydrides, proposing new compositions, Li–Na–H and Th–Y–H, along with their structures.
These ML studies illustrate promising strategies for screening potential hydride superconductors across vast compositional space.

In this study, we introduce a ML-based screening method coarsely exploring superconducting ternary hydrides under high pressure, based on high $T_\mathrm{c}$ values from element features.
Curating a dataset of binary and ternary hydride entries, we construct regression models to predict \(T_\mathrm{c}\) across a wide range of A–B–H compositions at high pressures (100, 200, and 300 GPa).
To improve predictive robustness, model outputs are evaluated statistically across independently trained ensemble members, with an emphasis on consistency among predictions rather than on single-point estimates.
The resulting scores constitute a composition-level metric derived from existing \(T_\mathrm{c}\) data and are used for primary screening, rather than for phase-resolved validation.

\section{Method}\label{sec:method}

\subsection{Dataset curation of hydride superconductors}

\begin{figure*}
    \centering
    \includegraphics{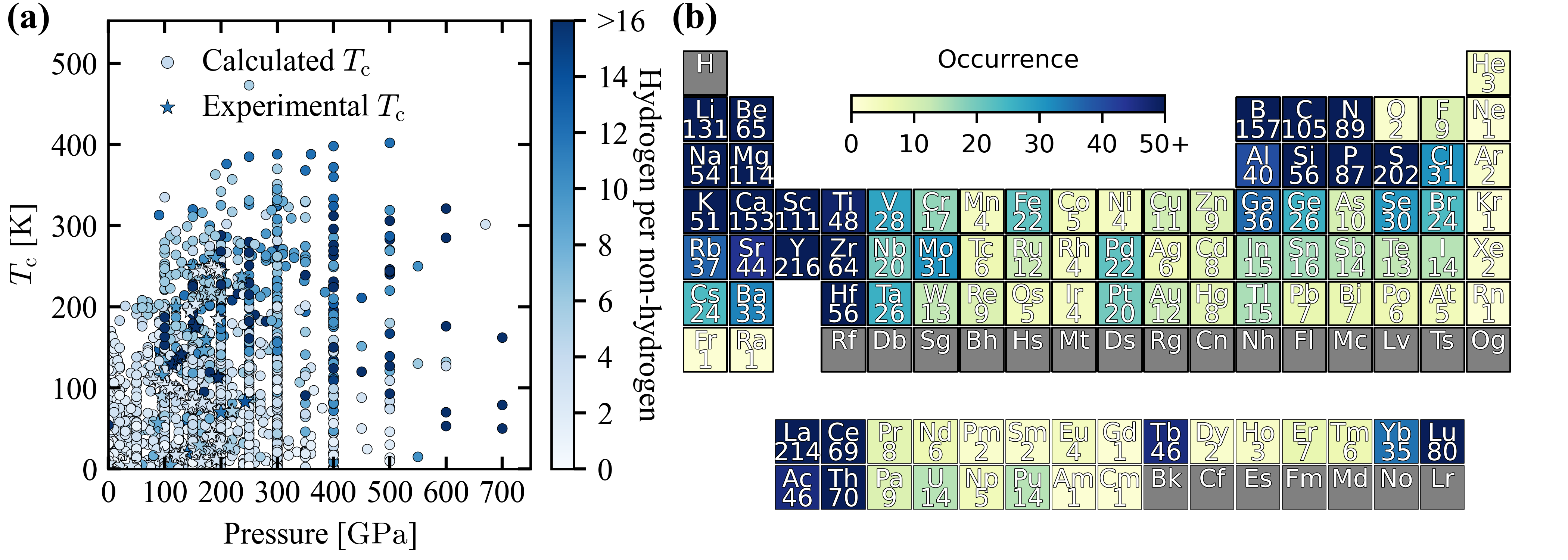}
    \caption{
Overview of the curated dataset of hydride superconductors used for model training.  
(a) Distribution of $T_\mathrm{c}$ values plotted against pressure and hydrogen per non-hydrogen atomic ratio.  
Higher $T_\mathrm{c}$ values are generally observed in hydrogen-rich compositions, particularly under high pressure.  
(b) Elemental occurrence heatmap showing the frequency of each element in the 2059 binary and ternary hydride compounds.  
Light elements, alkali metals, and alkaline earth metals are frequently represented, reflecting known experimental and computational focus areas.
    }
    \label{fig:dataset}
\end{figure*}

We curated a dataset of hydrides reported between 2007 and 2024 provided in the Supplementary Material, including their chemical formulas, applied pressures, and  \(T_\mathrm{c}\). 
The resulting dataset comprises 2059 records for binary and ternary hydride superconductors.
Figure~\ref{fig:dataset} shows (a) the distribution of \(T_\mathrm{c}\) values with respect to pressure and hydrogen-to-non-hydrogen ratio and (b) the elemental occurrence frequency in the hydride compositions.
In Fig.~\ref{fig:dataset}a, hydrogen-rich compositions tend to exhibit high $T_\mathrm{c}$.
From Fig.~\ref{fig:dataset}b, we can confirm that the investigated compounds comprise light elements, alkali metals, alkaline earth metals, and early transition metals.
The dataset is separated into 938 binary and 1121 ternary hydrides, and also into 74 experimentally determined values, 727 calculated using the Eliashberg theory\cite{eliashberg1960interactions}, and 1258 estimated with the Allen–Dynes formula\cite{AllenDynes1975-zz}.
We adopted the experimental, Eliashberg, and Allen–Dynes values in that order of priority when multiple \(T_\mathrm{c}\) values were reported for the same composition.
In cases where the theoretical estimations yielded a range of possible \(T_\mathrm{c}\) values depending on the chosen value of the empirical Coulomb parameter, the maximum value was adopted to mitigate the tendency of hydrides.

\begin{table}[htbp]
  \caption{
List of 22 element-specific physicochemical descriptors used to construct features for \(T_\mathrm{c}\) prediction.  
These descriptors were assigned to each atomic species in a compound, and used to compute compositional features (e.g., weighted averages, extrema) across all constituent elements.  
Descriptors were collected from the \textsf{pymatgen} and \textsf{Mendeleev} libraries.
  }
  \centering
  \begin{ruledtabular}
  \begin{tabular}{ll}
    Descriptor & Unit\\ \hline
    Atomic number                                           & -- \\
    Atomic mass                                             & u  \\
    Electron affinity (EA)                                  & eV \\
    Ionization energy (IE)                                  & eV \\
    Atomic radius                                           & pm \\
    van der Waals radius                                    & pm \\
    Covalent radius                                         & pm \\
    Mendeleev number                                        & -- \\
    Valence‑electron count                                  & -- \\
    Outer‑shell electron count                              & -- \\
    Polarizability                                          & bohr$^{3}$ \\
    Clementi effective nuclear charge                      & -- \\
    Number of \textit{s}‑electrons                                 & -- \\
    Number of \textit{p}‑electrons                                 & -- \\
    Number of \textit{d}‑electrons                                 & -- \\
    Number of \textit{f}‑electrons                                 & -- \\
    Allred–Rochow electronegativity                         & e$^{2}$/pm$^{2}$ \\
    Gordy electronegativity                                 & e/pm \\
    Mulliken electronegativity                              & eV \\
    Martynov–Batsan electronegativity                       & eV$^{0.5}$ \\
    Sanderson electronegativity                             & -- \\
    Nagle electronegativity                                 & bohr$^{-1}$ \\ 
  \end{tabular}
  \end{ruledtabular}
  \label{tab:sheet_of_descriptors_in_dataset}
\end{table}

We computed physicochemical descriptors for each chemical formula based on its atomic composition.
Table~\ref{tab:sheet_of_descriptors_in_dataset} enumerates 22 element-specific descriptors curated from the \textsf{pymatgen}\cite{Ong2013-ki} and \textsf{Mendeleev}\cite{mendeleev2014} libraries.
Missing Clementi effective nuclear charges for atomic numbers greater than 87 were linearly extrapolated from their Slater values.
For each descriptor, two representations were formulated: raw elemental values and atomic fraction-weighted values derived from the chemical formulas.
For both forms, five compositional statistics were computed across all constituent elements: sum, mean, range, maximum, and minimum.
Incorporating the synthesis or measurement pressure reported for each sample into the 220 descriptors resulted in a total of 221 features for predicting \(T_\mathrm{c}\).

\subsection{Feature engineering and model development}

As a preprocessing step for model training, we apply standardization and subsequently perform a feature filtering procedure. 
To distill the initial set of 221 features into a compact and informative subset, we employ a two-step filtering pipeline based on feature variance \(\mathrm{V}\) and the absolute value of the correlation coefficient \(|\mathrm{R}|\).
First, features with zero empirical variance across the dataset \(\mathrm{V}=0\) are removed.
Second redundant features are pruned based on correlation and mutual information: for each pair of highly correlated features (\(|\mathrm{R}|>|\mathrm{R}|_\mathrm{f}\)), the feature with the lower mutual information with respect to the target variable \(T_\mathrm{c}\) is discarded.
The correlation threshold \(|\mathrm{R}|_{\mathrm{f}}\) is selected to balance model accuracy and feature compactness through the protocol explained below.

Specifically, for each candidate threshold, we perform a Bayesian optimization with five-fold cross-validation (CV) to tune the regularization strength of Ridge regression, using implementations in \textsf{scikit-learn} and \textsf{skopt} packages.
Ridge regression is adopted here as a lightweight model that is less sensitive to multicollinearity, making it suitable for scanning \(|\mathrm{R}|_\mathrm{f}\).
The optimized Ridge model is then retrained on the entire training dataset, and the optimal threshold is selected based information criterion (IC) scores.
To balance the predictive focus of Akaike IC (AIC) and the parsimony favored Bayesian IC (BIC), we use their arithmetic mean as a pragmatic compromise.

For model optimization following feature filtering, we apply the same Bayesian optimization procedure with five-fold CV.
The following regression models are considered: Ridge linear regression (Ridge), support vector regression (SVR), decision tree (DT), random forest (RF), gradient boosting (GB), and XGBoost (XGB)\cite{XGBoost2016-zz}. 
Their model performance is evaluated using a nested CV protocol with five outer folds. 
In each outer fold, models are trained using the preprocessing and the optimization procedure through five inner folds.
The best performing model is selected based on the average scores across outer folds, considering root mean squared error (RMSE), mean absolute error (MAE), and coefficient of determination (R$^2$).
Because the model relies exclusively on composition-based descriptors, entries sharing identical or closely related compositions may appear across training and test splits.
Accordingly, the resulting performance metrics should be interpreted as reflecting the model’s ability to capture statistically robust composition-level trends within the explored chemical space, rather than its capacity for strict extrapolation to completely unseen chemistries
To ensure robustness in downstream applications, multiple instances of the selected model architecture are trained with each different random seed.

\subsection{Screening ternary hydride superconductors}

We introduce a systematic screening procedure to investigate ternary hydrides with high \(T_\mathrm{c}\).  
The search space comprises all A--B--H compositions, where elements A and B span atomic numbers \(2 \leq Z \leq 90\), resulting in 3916 distinct elemental pairs.  
For each pair, 4851 compositional grid points are generated by varying atomic fractions within the ternary composition triangle in 1\% increments.  
This results in over 18 million unique ternary compositions evaluated in total.  
At each grid point, an ensemble of 30 independently trained models predicts \(T_\mathrm{c}\), from which the mean and the 95\% confidence interval (CI) are computed, assuming a \(t\)-distribution of ensemble predictions.  
Promising A--B pairs are identified by ranking the compositions according to the lower bound of the predicted 95\% CI.  
This screening is independently performed at 100, 200, and 300~GPa to span experimentally relevant pressure ranges (see Fig.~\ref{fig:dataset}a).

\section{Results and Discussions}\label{sec:results}

\subsection{Selection of regression model}

To select a regression model for subsequent high-throughput screening, we evaluated the predictive performance of several algorithms using a nested CV protocol with five outer folds.  
Each outer fold included an inner five-fold CV loop for hyperparameter tuning and feature selection via correlation-based filtering.  
The results, summarized in Table~\ref{tab:ScoresOfModels}, report the RMSE, MAE, and R\(^2\) for each tested model.

\begin{table}[htbp]
  \caption{
Predictive performance of candidate regression models for \(T_\mathrm{c}\) estimation, evaluated by nested cross-validation (5 outer \(\times\) 5 inner folds).  
Reported metrics include the RMSE, MAE, and R\(^2\).  
The XGB model was selected for downstream screening based on its superior performance.
  }
  \centering
  \begin{ruledtabular}
  \begin{tabular}{l l l l}
    Model\makebox[0.5cm]{}  &RMSE [K]\makebox[0.5cm]{}  & MAE [K]\makebox[0.5cm]{}   & R\(^2\)   \\ \hline
    Ridge	   &64.1 \(\pm\) 2.4 & 49.0 \(\pm\) 2.0 & 0.45 \(\pm\) 0.03 \\ 
    SVR	     &47.3 \(\pm\) 2.7 & 31.4 \(\pm\) 1.8 & 0.70 \(\pm\) 0.05 \\ 
    DT	     &50.8 \(\pm\) 4.2 & 32.0 \(\pm\) 2.4 & 0.65 \(\pm\) 0.06 \\
    RF	     &40.8 \(\pm\) 3.5 & 27.6 \(\pm\) 1.8 & 0.78 \(\pm\) 0.05 \\
    GB	     &39.2 \(\pm\) 3.0 & 25.4 \(\pm\) 1.9 & 0.79 \(\pm\) 0.04 \\ 
    XGB	     &38.5 \(\pm\) 3.8 & 24.7 \(\pm\) 2.2 & 0.80 \(\pm\) 0.04 \\ 
  \end{tabular}
  \end{ruledtabular}
  \label{tab:ScoresOfModels}
\end{table}

Among the tested models, XGB achieved the best performance, with the lowest RMSE of \SI{38.5 \pm 3.8}{K}, the lowest MAE of \SI{24.7 \pm 2.2}{K}, and the highest coefficient of determination, R\(^2 = 0.80 \pm 0.04\).  
While the GB performed comparably within statistical uncertainty, we selected XGB due to its superior regularization scheme and practical advantages for ensemble training.  
Both GB and XGB are ensemble methods based on gradient-boosted decision trees; hence, their performance proximity is expected.  
Given its favorable accuracy, robustness, and scalability, we adopted XGB as the core model for downstream screening of ternary hydride superconductors.  
We note that, since the hydride dataset contains higher $T_\mathrm{c}$ values, our model yields a relatively larger RMSE (38.5 K) compared to the prior model for binary hydrides (33.7 K)\cite{Hutcheon2020-dk}, and a lower value of R$^2( = 0.80)$ compared to that reported for the SuperCon dataset (R$^2 = 0.92$)\cite{Konno2021-vg}.
The hyperparameter search space for Bayesian optimization used in this comparison is detailed in Table~S1.

\subsection{Training and feature importance of an XGBoost ensemble}

Following the model selection described above, we examined the statistical properties of the trained XGB models to assess the robustness and interpretability of the ensemble used for screening.
To ensure robustness in downstream screening, we trained 30 independent XGB models, each using a different random seed but the same curated dataset.  
During feature filtering, the resulting optimal correlation thresholds \(|\mathrm{R}|_\mathrm{f}\) were narrowly distributed (\(0.86 \pm 0.01\)), with the number of retained features also showing little variation (\(68.7 \pm 3.1\)).  
Each XGB model was subsequently trained on the corresponding filtered feature set.

Although some of the optimized hyperparameters varied across repeated runs, the prediction performance remained stable, with a training RMSE of \SI{9.8 \pm 2.9}{K}, indicating that the protocol yields robust models despite the non-uniqueness of the optimal configuration (see Figs.~S1 and S2).  
This stability suggests that the model ensemble captures global optima reliably. 
Moreover, the observed insensitivity of performance across several orders of magnitude in hyperparameter space implies that the search range could be narrowed in future applications to enhance efficiency.


\begin{figure}[htbp]
    \centering
    \includegraphics{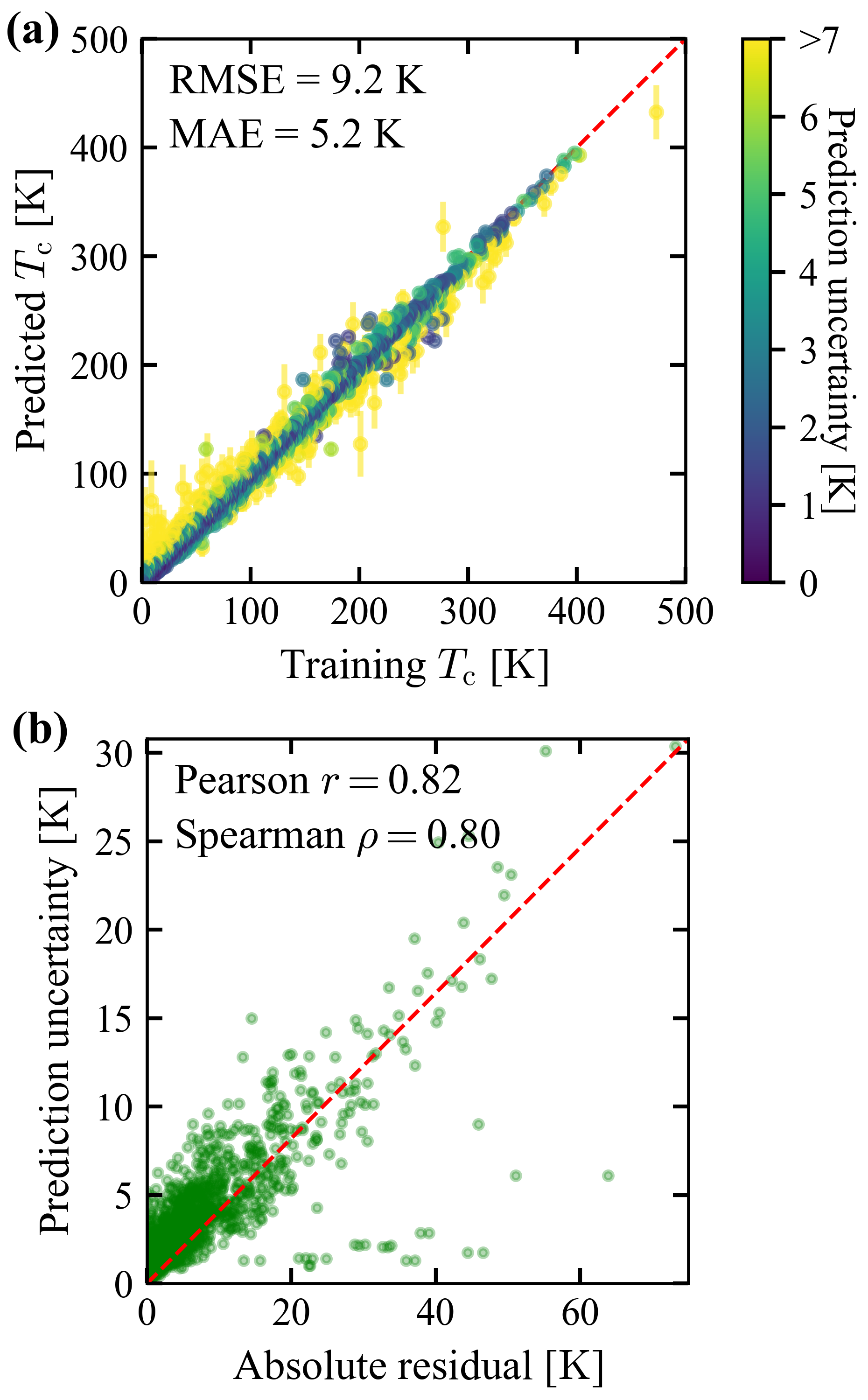}
    \caption{
Evaluation of predictive fidelity and uncertainty estimation across 30 trained XGB models.  
(a) Parity plot comparing predicted and training \(T_\mathrm{c}\) values, where error bars and color saturation indicate the standard deviation across ensemble predictions.   
(b) Correlation between prediction uncertainty (ensemble standard deviation) and absolute residuals, showing strong agreement with Pearson \(r = 0.82\) and Spearman \(\rho = 0.80\).  
The red dashed line shows the best-fit linear regression with a slope of 0.411, supporting the use of predicted variance as a confidence proxy.
    }
    \label{fig:parity}
\end{figure}

The statistical characteristics of the 30 trained XGB models are analyzed to evaluate both prediction accuracy and the validity of the uncertainty estimates.  
Figure~\ref{fig:parity}a shows the parity between predicted and training \(T_\mathrm{c}\) values, where each point represents a training data sample, with error bars and color indicating the standard deviation across the 30 models.  
The ensemble mean predictions achieve a training RMSE of \SI{9.2}{K} and an MAE of \SI{5.2}{K}, demonstrating that the models successfully reproduce the overall trend in the training data.

To assess whether the predicted uncertainty reflects actual model confidence, we analyzed the relationship between the standard deviation of ensemble predictions and the absolute residuals from the true values.  
Figure~\ref{fig:parity}b demonstrates a positive correlation (Pearson \(r = 0.82\), Spearman \(\rho = 0.80\)), indicating that larger predicted uncertainties are associated with larger prediction errors.  
This observation supports the interpretation of the ensemble standard deviation as a meaningful proxy for predictive confidence, and justifies its use for computing confidence intervals in subsequent screening.

\begin{figure}[htbp]
    \centering
    \includegraphics{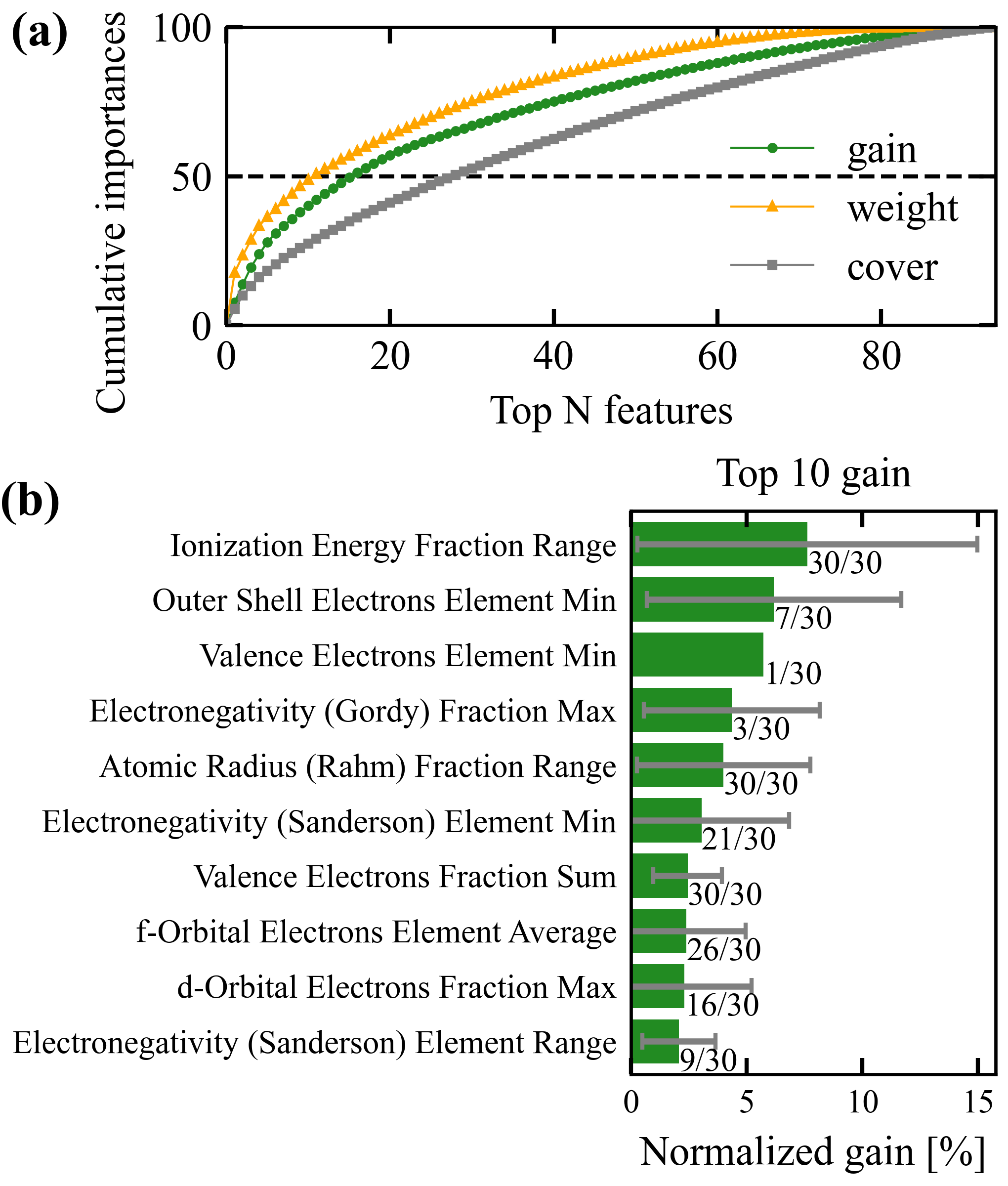}
    \caption{
Feature importance analysis across 30 trained XGB models to assess interpretability and predictor diversity.  
(a) Cumulative importance curves based on gain, weight, and cover metrics aggregated over the ensemble.  
The gain curve shows a moderately distributed profile, with the top 16 features accounting for 50\% of total importance, indicating no single dominant variable.  
(b) Top 10 features ranked by mean normalized gain.  
Error bars represent the standard deviation across models, and the label \(N\)/30 indicates how many models selected each feature.  
Commonly selected descriptors such as ionization energy and atomic radius suggest physically meaningful contributors to the predicted \(T_\mathrm{c}\).
    }
    \label{fig:feature-importance}
\end{figure}

To complement this analysis, we examined the interpretability of the trained models by analyzing feature importance across the 30 independently trained XGB instances.
Feature importance was evaluated using three commonly used metrics in XGBoost: gain (improvement in loss), weight (frequency of feature usage), and cover (number of observations impacted).  
Figure~\ref{fig:feature-importance}a shows cumulative importance curves ranked by each metric across the 30 trained models.  
The gain-based curve, in particular, exhibits a moderately distributed profile, with 50\% of the total importance covered by the top 16 features.  
This indicates that model predictions rely on a relatively broad set of descriptors rather than a small dominant subset.

Figure~\ref{fig:feature-importance}b highlights the top 10 features ranked by the mean normalized gain across the ensemble.  
Although the gain values exhibit non-negligible variation across models, as reflected in the error bars, several features, including ionization energy, atomic radius, and valence electron count, were consistently selected (appearing in most of the 30 models).  
These features are chemically intuitive and align with expected contributors to bonding strength and electronic structure in hydrogen-rich systems.  
Overall, the results suggest that the model leverages multiple moderately important descriptors, consistent with the multivariate nature of \(T_\mathrm{c}\) in complex materials.

\subsection{Screening results of hydride superconductors}

We performed a systematic screening of A--B--H compositions using the ensemble of 30 independently trained XGB models.  
To prioritize robust predictions, we evaluated each composition based on the lower bound of the 95\% CI of the predicted mean superconducting transition temperature (\(T_\mathrm{c}\)).  
The search space consisted of 3916 distinct A--B elemental pairs, each evaluated over 4851 grid points uniformly distributed within the ternary composition triangle.

\begin{figure}[htbp]
    \centering
    \includegraphics[width=1.0\linewidth]{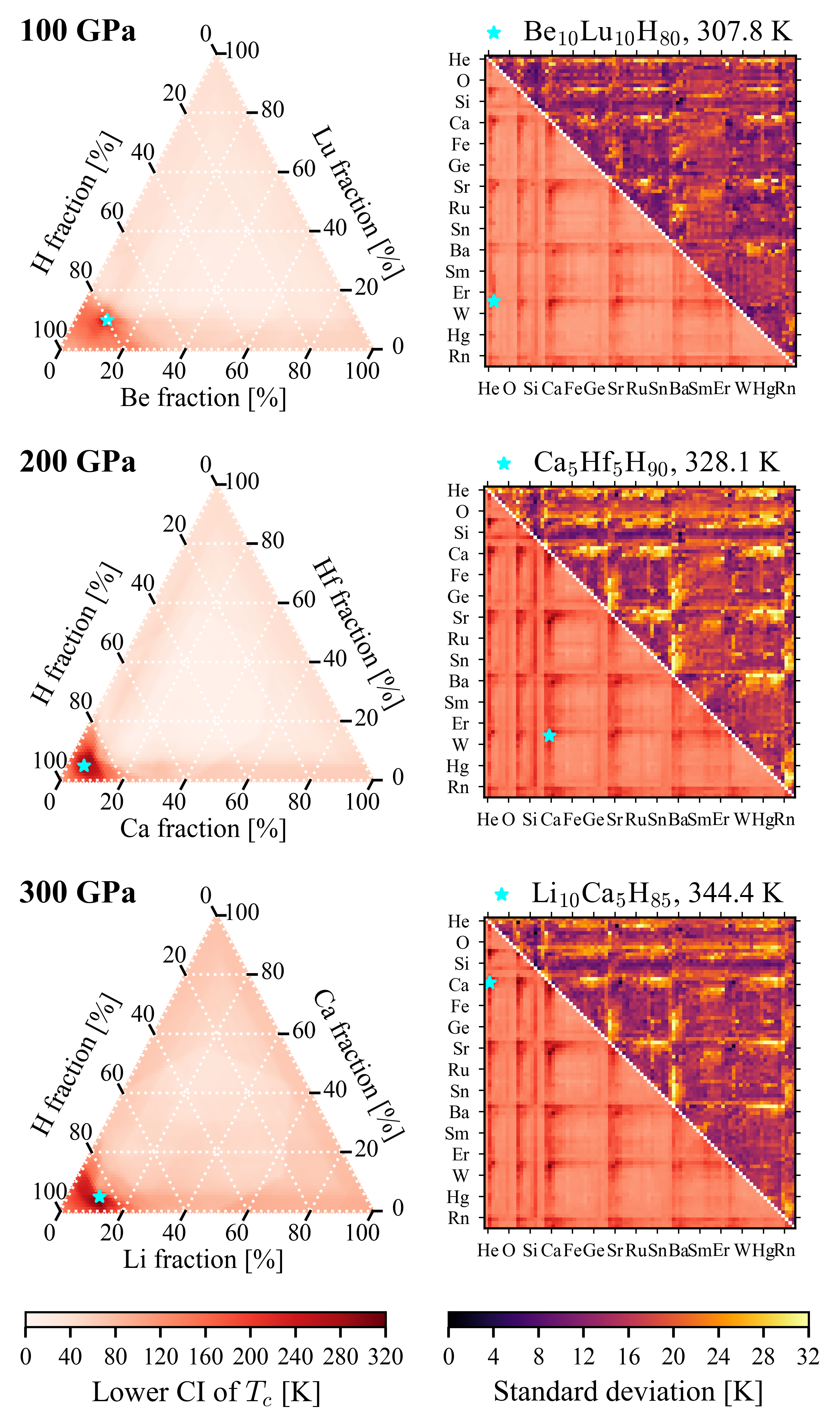}
    \caption{
Screening results for high-\(T_\mathrm{c}\) ternary hydrides at 100, 200, and 300~GPa.  
Left panels: Ternary composition maps showing the lower bound of the 95\% CI of predicted \(T_\mathrm{c}\) values across the A--B--H space.  
Cyan stars indicate the highest-ranked compositions at each pressure, annotated with their formula and predicted lower CI value.  
Right panels: Heatmaps for each A--B elemental pair, with lower triangles showing the predicted 95\% CI lower bound of \(T_\mathrm{c}\) and upper triangles showing the standard deviation of ensemble predictions.  
These maps capture both performance and model uncertainty across the elemental search space.  
All top candidates are located in hydrogen-rich regions and exhibit high \(T_\mathrm{c}\) with low uncertainty, highlighting their potential as robust superconductors.
    }
    \label{fig:search-space}
\end{figure}

Figure~\ref{fig:search-space} presents the prediction landscape at 100, 200, and 300~GPa, providing a compositional map of the most promising ternary hydride superconductors.  
Left panels show ternary plots of the lower bound of the 95\% CI across A--B--H compositions.  
Right panels display heatmaps for each A--B pair, showing the predicted mean standard deviation (upper triangles) and lower CI values (lower triangles), which capture both uncertainty and performance.  
The top candidate at each pressure is marked by a cyan star, annotated with its formula and predicted lower bound of \(T_\mathrm{c}\):  
Be$_{10}$Lu$_{10}$H$_{80}$ (\SI{307.8}{K} at 100~GPa),  
Ca$_5$Hf$_5$H$_{90}$ (\SI{328.1}{K} at 200~GPa),  
and Li$_{10}$Ca$_{5}$H$_{85}$ (\SI{344.4}{K} at 300~GPa).

\begin{table}[htbp]
\centering
\caption{
Top 10 predicted ternary hydride superconductors at 100, 200, and 300~GPa, ranked by the lower bound of the 95\% CI of the predicted \(T_\mathrm{c}\).  
Each row lists the element pair (A, B), whether that pair was included in the training data (``Trained''), the optimal stoichiometry, and the lower CI bound of \(T_\mathrm{c}\).  
These candidates were selected from over 18 million screened compositions.  
The ``Trained'' column highlights extrapolative predictions, including systems such as Ca--Ti and Li--K, which were absent from the training set.  
Full candidate rankings are provided in Supplementary Tables S2--S4.
}
\begin{ruledtabular}
\begin{tabular}{lccc ccc}
P [GPa]& A & B & Trained & Best formula & Lower CI of \(T_\mathrm{c}\) [K] \\ \hline
100 & Be & Lu & Yes & Be$_{10}$Lu$_{10}$H$_{80}$ & 307.8\\
& Ca & Hf & Yes & Ca$_{7}$Hf$_{7}$H$_{86}$ & 265.4\\
& Sc & Lu & Yes & Sc$_{4}$Lu$_{10}$H$_{86}$ & 263.1\\
& Ca & Lu & Yes & Ca$_{5}$Lu$_{9}$H$_{86}$ & 261.4\\
& Ca & Zr & Yes & Ca$_{7}$Zr$_{7}$H$_{86}$ & 258.1\\
& Be & Ca & Yes & Be$_{10}$Ca$_{10}$H$_{80}$ & 256.2\\
& Y & Lu & Yes & Y$_{3}$Lu$_{11}$H$_{86}$ & 255.1\\
& Ca & Ti & No & Ca$_{5}$Ti$_{5}$H$_{90}$ & 233.1\\
& Sc & La & Yes & Sc$_{8}$La$_{3}$H$_{89}$ & 229.9\\
& Li & Na & Yes & Li$_{10}$Na$_{5}$H$_{85}$ & 229.3\\ 
\hline 200 
& Ca & Hf & Yes & Ca$_{5}$Hf$_{5}$H$_{90}$  & 328.1\\
& Ca & Zr & Yes & Ca$_{5}$Zr$_{5}$H$_{90}$  & 324.9\\
& Li & Na & Yes & Li$_{10}$Na$_{5}$H$_{85}$ & 314.9\\
& Li & Mg & Yes & Li$_{11}$Mg$_{5}$H$_{84}$ & 313.1\\
& Na & Mg & No & Na$_{10}$Mg$_{5}$H$_{85}$ & 301.6\\
& Ca & Ti & No & Ca$_{5}$Ti$_{5}$H$_{90}$ & 298.4\\
& Li & Ca & Yes & Li$_{11}$Ca$_{4}$H$_{85}$ & 298.2\\
& Sc & La & Yes & Sc$_{8}$La$_{3}$H$_{89}$ & 292.4\\
& Sc & Y & Yes & Sc$_{8}$Y$_{3}$H$_{89}$ & 290.5\\
& Ca & Sc & Yes & Ca$_{5}$Sc$_{5}$H$_{90}$ & 288.2\\ 
\hline
300 & Li & Ca & Yes & Li$_{10}$Ca$_{5}$H$_{85}$  & 344.4\\
& Ca & Hf & Yes & Ca$_{5}$Hf$_{5}$H$_{90}$ & 340.7\\
& Li & Na & Yes & Li$_{10}$Na$_{5}$H$_{85}$ & 339.1\\
& Ca & Zr & Yes & Ca$_{5}$Zr$_{5}$H$_{90}$ & 331.8\\
& Na & Ca & No & Na$_{10}$Ca$_{5}$H$_{85}$ & 313.9\\
& Li & Mg & Yes & Li$_{11}$Mg$_{4}$H$_{85}$ & 305.4\\
& Na & Mg & No & Na$_{10}$Mg$_{5}$H$_{85}$ & 303.8\\
& Sc & Y & Yes & Sc$_{8}$Y$_{3}$H$_{89}$ & 296.3\\
& Ca & Ti & No & Ca$_{5}$Ti$_{5}$H$_{90}$ & 295.3\\
& Li & K & No & Li$_{11}$K$_{4}$H$_{85}$ & 292.4\\
\end{tabular}
\end{ruledtabular}
\label{tab:top10}
\end{table}
\begin{table}[htbp]

\centering
\caption{
Top 10 predicted superconducting ternary hydrides absent from the training dataset, selected by the lower bound of the 95\% confidence interval of \(T_\mathrm{c}\).}
\begin{ruledtabular}
\begin{tabular}{lccc ccc}
P [GPa]& A & B & Trained & Best formula & Lower CI of \(T_\mathrm{c}\) [K] \\ \hline
100 
&Ca & Ti & No & Ca$_{5}$Ti$_{5}$H$_{90}$ & 233.1\\
&Zr & Lu & No & Zr$_{4}$Lu$_{10}$H$_{86}$ & 221.5\\
&Na & Mg & No & Na$_{11}$Mg$_{4}$H$_{85}$ & 218.4\\
&Ti & Lu & No & Ti$_{4}$Lu$_{10}$H$_{86}$ & 217.7\\
&Ca & Nb & No & Ca$_{11}$Nb$_{4}$H$_{85}$ & 216.1\\
&Ca & V & No & Ca$_{11}$V$_{4}$H$_{85}$ & 215.1\\
&Lu & Hf & No & Lu$_{10}$Hf$_{4}$H$_{86}$ & 212.4\\
&Ti & Sr & No & Ti$_{5}$Sr$_{5}$H$_{90}$ & 211.2\\
&Ca & Ta & No & Ca$_{10}$Ta$_{4}$H$_{86}$ & 209.4\\
&Sc & Ac & No & Sc$_{4}$Ac$_{10}$H$_{86}$ & 209.0\\
\hline 200 
&Na & Mg & No & Na$_{10}$Mg$_{5}$H$_{85}$ & 301.6\\
&Ca & Ti & No & Ca$_{5}$Ti$_{5}$H$_{90}$ & 298.4\\
&Na & Ca & No & Na$_{10}$Ca$_{5}$H$_{85}$ & 280.7\\
&Li & K & No & Li$_{11}$K$_{4}$H$_{85}$ & 276.6\\
&Ti & Sr & No & Ti$_{5}$Sr$_{5}$H$_{90}$ & 272.1\\
&Mg & K & No & Mg$_{5}$K$_{10}$H$_{85}$ & 269.7\\
&Sc & Sr & No & Sc$_{6}$Sr$_{4}$H$_{90}$ & 266.2\\
&Ca & V & No & Ca$_{7}$V$_{3}$H$_{90}$ & 260.5\\
&Li & Rb & No & Li$_{12}$Rb$_{3}$H$_{85}$ & 259.7\\
&Sc & Ac & No & Sc$_{8}$Ac$_{3}$H$_{89}$ & 258.6\\
\hline 300 
&Na & Ca & No & Na$_{10}$Ca$_{5}$H$_{85}$ & 313.9\\
&Na & Mg & No & Na$_{10}$Mg$_{5}$H$_{85}$ & 303.8\\
&Ca & Ti & No & Ca$_{5}$Ti$_{5}$H$_{90}$ & 295.3\\
&Li & K & No & Li$_{11}$K$_{4}$H$_{85}$ & 292.4\\
&Mg & K & No & Mg$_{5}$K$_{10}$H$_{85}$ & 277.0\\
&Na & K & No & Na$_{10}$K$_{4}$H$_{86}$ & 276.6\\
&Li & Rb & No & Li$_{12}$Rb$_{3}$H$_{85}$ & 271.5\\
&K & Ca & No & K$_{10}$Ca$_{5}$H$_{85}$ & 270.8\\
&Ti & Sr & No & Ti$_{5}$Sr$_{5}$H$_{90}$ & 263.3\\
&Na & Rb & No & Na$_{10}$Rb$_{3}$H$_{87}$ & 262.9\\
\end{tabular}
\end{ruledtabular}
\label{tab:top10none}
\end{table}

Table~\ref{tab:top10} lists the top 10 candidate hydrides at each pressure condition, ranked by the lower bound of the predicted 95\% CI of \(T_\mathrm{c}\).  
Consistently across all pressures, the top candidates exhibit extremely hydrogen-rich stoichiometries.  
To further investigate the extrapolative capabilities of the model, we extracted the top-ranked superconducting ternary hydrides whose A–B elemental pairs were not included in the training dataset.
Table~\ref{tab:top10none} lists the top 10 such compositions at each pressure condition (100, 200, and 300~GPa), ranked by the lower bound of the predicted 95\% confidence interval of \(T_\mathrm{c}\).

Alkali and alkaline earth metals appear frequently, reproducing the elemental trends observed in the training dataset (see Fig.~\ref{fig:dataset}).  
Element pairs such as Ca--Hf, Ca--Zr, and Li--Na recur across multiple pressure conditions, indicating compositional robustness and favorable bonding environments.

Importantly, several element pairs such as Ca--Ti, Na--Mg, and Li--K emerge as top candidates despite being absent from the training data.  
Among these, the Ca--Ti system is particularly notable: it appears in the top 10 at all three pressure conditions, with lower CI values exceeding \SI{230}{K} at 100~GPa and surpassing \SI{295}{K} at 300~GPa.  
From a chemical perspective, Ca and Ti are alkaline earth and early transition metals, respectively. 
The predicted composition Ca$_5$Ti$_5$H$_{90}$ exhibits an exceptionally high hydrogen content.  
The Ca--Ti system combines elements with a high atomic radius range and valence electron counts.  
Such a stoichiometry is expected to promote conventional phonon-mediated superconductivity.  
These findings highlight the potential of the system beyond known chemistries and identify novel superconducting materials.
In future work, the individual stabilities of the screened hydrides should be verified by DFT calculations.

\section{Conclusion}\label{sec:conclusion}

In this study, we developed a ML-based framework to explore chemical compositions of superconducting ternary hydride systems under high pressure.
We curated a dataset of 2059 binary and ternary hydride superconductors and trained an ensemble of 30 independently initialized XGBoost regression models to predict the superconducting transition temperature.
The models were applied to millions of A–B–H compositions across 100, 200, and 300 GPa, and ensemble statistics were used to evaluate prediction consistency.
We also analyzed feature importance across models to examine the physicochemical descriptors contributing to prediction performance.

Our screening results identified several promising candidates such as Ca–Ti–H, Li–K–H, and Na–Mg–H that were not present in the training data.
Feature analysis revealed that descriptors like ionization energy, atomic radius, and valence electron count consistently contribute to predictions, aligning with chemical intuition about bonding and electronic structure in hydrogen-rich environments.
These findings validate the effectiveness of our approach for hydride screening and feature analysis.

While the present study does not explicitly incorporate structural symmetries or their associated stability and dynamical effects, as examined in Refs.\cite{Wang2025-jj,Belli2025-tx}, this choice is intentional rather than a limitation of the methodology.
The primary objective of this work is to enable rapid, composition-level screening of vast chemical spaces using existing superconducting transition temperature data, at a stage where reliable phase-resolved structural information is often unavailable.
Meaningful validation based on density-functional perturbation theory requires the identification of stable crystal structures and convergence-controlled phonon and electron–phonon coupling calculations, which constitute a distinct and substantially more resource-intensive research task.

Within this intended scope, the results demonstrate the utility of ensemble-based statistical learning as a primary screening tool for identifying high-scoring compositional regions in chemically complex superconducting systems.
Future work may integrate stability filters or couple the present framework with crystal structure prediction methods to refine candidate selection once specific compositions of interest have been identified.
In addition, further studies aimed at elucidating the physical origins of the learned composition-level trends, for example through explainable artificial intelligence techniques, will be valuable in bridging data-driven screening and microscopic theory\cite{Masuda2024,Masuda2025}.
Overall, the proposed framework provides a practical and scalable route for accelerating the exploration of high-\(T_\mathrm{c}\) superconductors across vast chemical spaces.

\section*{Supplementary Material}

We provide the following supplementary material: 
(1) curated dataset of binary and ternary hydride superconductors,
(2) hyperparameter search space for Bayesian optimization, 
(3) distributions of optimized hyperparameters across 30 XGB models, 
(4) distributions of RMSE, MAE, and R\(^2\) across 30 XGB models, 
and (5) full candidate rankings from the screening at 100, 200, and 300~GPa.
The dataset is entitled as ``dataset.csv'' and the candidate rankings as ``screening\_results\_100GPa.csv'', ``screening\_results\_200GPa.csv'', and ``screening\_results\_300GPa.csv''.

\begin{acknowledgments}
This study was financially supported, in part, by the grant number KSAC2025-14, the Japan Science and Technology Agency.
\end{acknowledgments}

\section*{Data Availability Statement}
The data that support the findings of this study are available from the corresponding author upon reasonable request.

\bibliographystyle{aipnum4-1} 
\bibliography{paper}

\end{document}


\title{Supplementary Material for: \\ ``Composition-Based Machine Learning for Screening Superconducting Ternary Hydrides from a Curated Dataset''}
\author{Kazuaki Tokuyama}
\affiliation{Department of Chemical Engineering, Kyoto University,
	Kyoto 615-8510, Japan}

\author{Souta Miyamoto}
\affiliation{Department of Chemical Engineering, Kyoto University,
	Kyoto 615-8510, Japan}
\affiliation{School of Chemical Science and Technology, Kyoto University, Nishikyo, Kyoto 615-8510, Japan}

\author{Taichi Masuda}
\affiliation{Department of Chemical Engineering, Kyoto University,
	Kyoto 615-8510, Japan}

\author{Katsuaki Tanabe}
\thanks{Corresponding author.\\
	\href{mailto:tanabe@cheme.kyoto-u.ac.jp}{tanabe@cheme.kyoto-u.ac.jp}}
\affiliation{Department of Chemical Engineering, Kyoto University,
	Kyoto 615-8510, Japan}
\affiliation{School of Chemical Science and Technology, Kyoto University, Nishikyo, Kyoto 615-8510, Japan}
\affiliation{Kyoto MPI Inc., Sakyo, Kyoto 606-8501, Japan}

\date{\today}

\maketitle

\section{Curated Dataset}
We attach the file \textsf{dataset.csv} that comprises our curated dataset.

\section{Hyperparameter search spaces of XGBoost model}

\begin{table}[H]
\centering
\caption{Hyperparameter search spaces for baseline regression models used in comparison. Log-uniform priors were used for parameters spanning several orders of magnitude.}
\begin{adjustbox}{max width=0.8\linewidth}
\begin{ruledtabular}
\begin{tabular}{llll}
Model & Parameter & Search Range & Scale\\ \hline
Ridge & \texttt{alpha} & [$10^{-4}$, $10^{2}$] & Log-uniform \\
\hline
SVR & \texttt{C} & [0.1, 1000] & Log-uniform\\
    & \texttt{epsilon} & [0.001, 100] & Log-uniform \\
    & \texttt{gamma} & [0.001, 1.0] & Log-uniform\\
\hline
DT & \texttt{max\_depth} & [1, 20] & Uniform \\
    & \texttt{min\_samples\_split} & [2, 20] & Uniform \\
    & \texttt{max\_features} & \{None, sqrt, log2, 1\} & Categorical \\
\hline
RF & \texttt{n\_estimators} & [10, 2000] & Uniform \\
    & \texttt{max\_depth} & [1, 10] & Uniform \\
    & \texttt{max\_features} & \{None, sqrt, log2, 1\} & Categorical \\
\hline
GB  & \texttt{learning\_rate} & [0.01, 0.5] & Log-uniform \\
    & \texttt{max\_depth} & [1, 10] & Uniform \\
    & \texttt{n\_estimators} & [10, 2000] & Uniform \\
    & \texttt{max\_features} & \{None, sqrt, log2, 1\} & Categorical\\
\hline
XGB & \texttt{learning\_rate}      & [0.01, 0.5]           & Log-uniform \\
    & \texttt{n\_estimators}       & [10, 2000]            & Integer \\
    & \texttt{max\_depth}          & [1, 10]               & Integer \\
    & \texttt{subsample}           & [0.1, 1.0]            & Uniform \\
    & \texttt{colsample\_bytree}   & [0.1, 1.0]            & Uniform \\
    & \texttt{gamma}               & [0.0, 5.0]            & Uniform \\
    & \texttt{reg\_alpha}          & [$10^{-4}$, $10^{2}$] & Log-uniform \\
    & \texttt{reg\_lambda}         & [$10^{-4}$, $10^{2}$] & Log-uniform \\
\end{tabular}
\end{ruledtabular}
\end{adjustbox}
\label{tab:param_space}
\end{table}

\begin{figure}[H]
    \centering
    \includegraphics[width=1.0\linewidth]{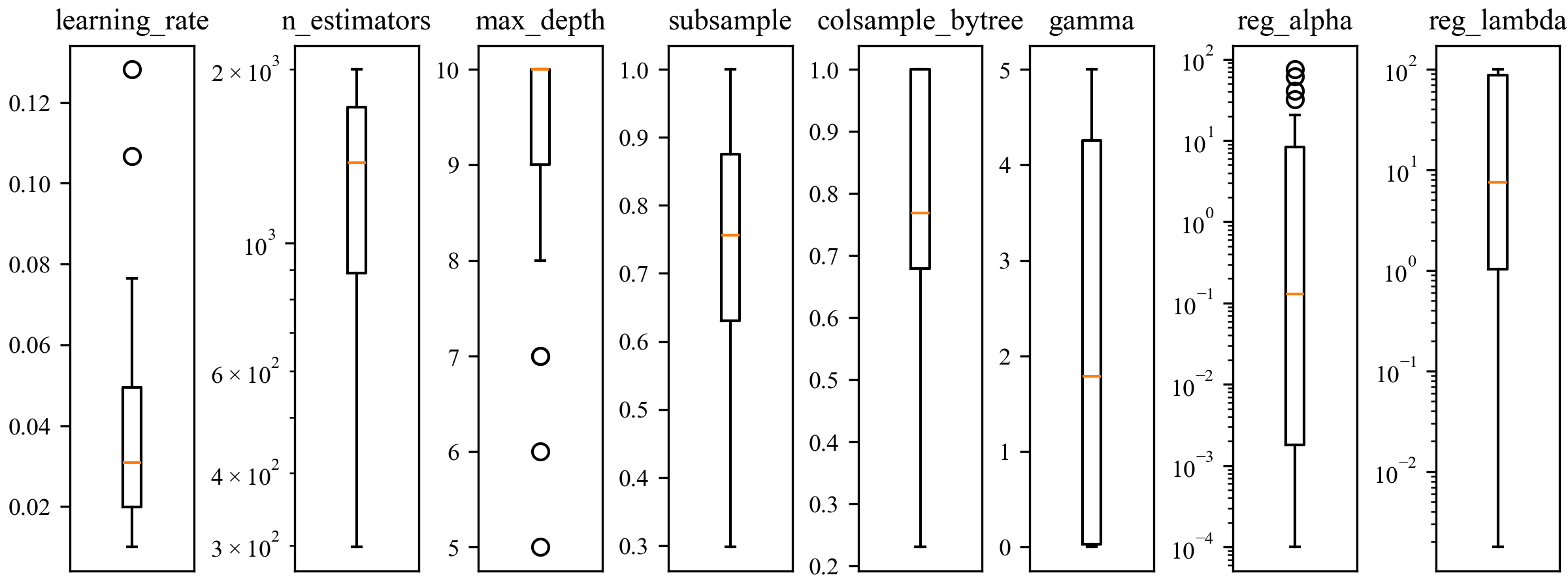}
    \caption{Boxplots of optimized hyperparameters for XGB instances across 30 optimization runs with different random seeds. Despite the variability in some hyperparameters, the resulting model performances remain stable across runs, indicating robustness to hyperparameter configuration (see Fig.~\ref{fig:score_boxplot}). The upper bound of \texttt{max\_depth} was fixed at 10 to prevent overfitting.}
    \label{fig:hypr_boxplot}
\end{figure}

\begin{figure}[H]
    \centering
    \includegraphics[width=0.45\linewidth]{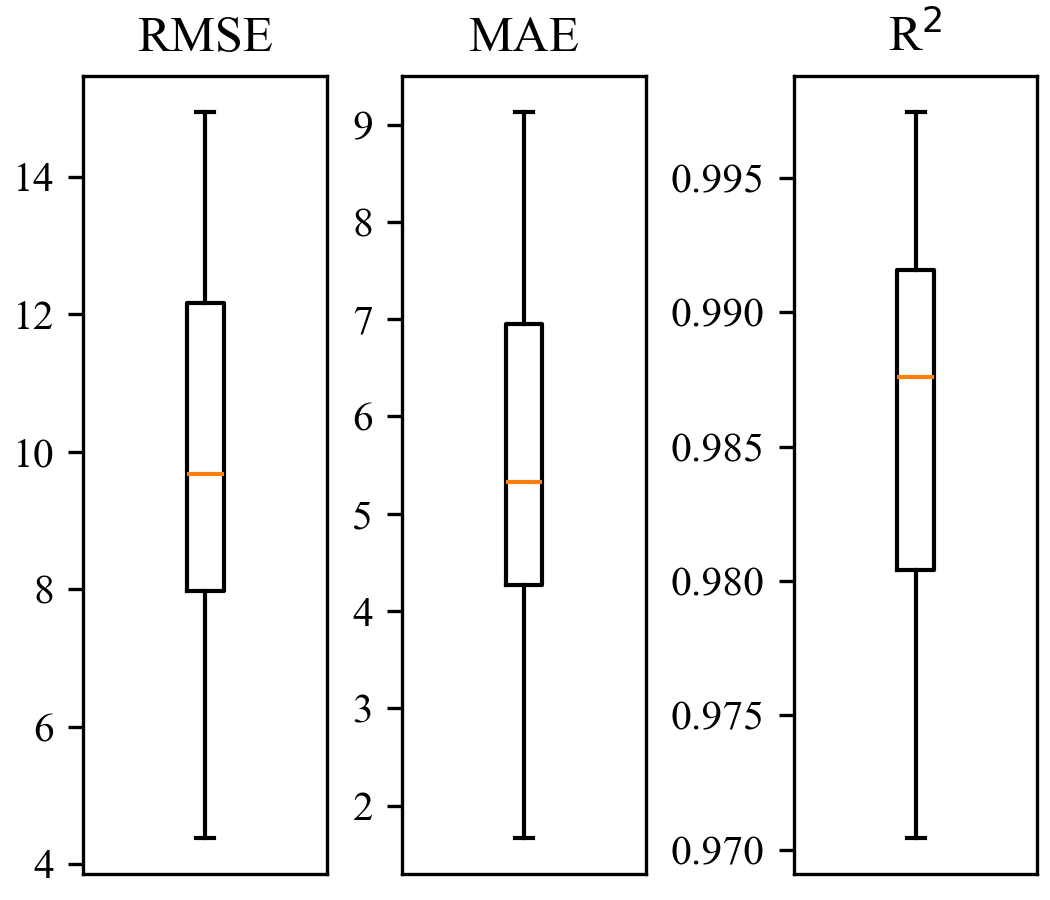}
    \caption{Boxplots of prediction metrics (RMSE, MAE, and R$^2$) over 30 XGBoost instances trained with different random seeds. The narrow spread of these metrics demonstrates consistent performance, suggesting that the model is robust with respect to initialization and minor variations in training.}
    \label{fig:score_boxplot}
\end{figure}

\section{Full screening results}

We attach three csv files listing the screened candidates:
\textsf{screening\_results\_100GPa.csv}, \textsf{screening\_results\_200GPa.csv}, and \textsf{screening\_results\_300GPa.csv}.